\documentclass[aps, prb, twocolumn, amssymb, amsmath, showpacs, superscriptaddress]{revtex4-1}

\usepackage{bm}
\usepackage{times}
\usepackage{graphicx}
\usepackage{dcolumn}
\usepackage{graphicx}

\begin{document}

\title{Large magneto-optical Kerr effect in noncollinear antiferromagnets Mn$_{3}X$ ($X$ = Rh, Ir, or Pt)}

\author{Wanxiang Feng}
\affiliation {Department of Physics, National Taiwan University, Taipei 10617, Taiwan}
\affiliation {School of Physics, Beijing Institute of Technology, Beijing 100081, China}
%\affiliation {Institute of Physics, Academia Sinica, Taipei 11529, Taiwan}

\author{Guang-Yu Guo}
\email{gyguo@phys.ntu.edu.tw}
\affiliation {Department of Physics, National Taiwan University, Taipei 10617, Taiwan}
\affiliation {Physics Division, National Center for Theoretical Sciences, Hsinchu 30013, Taiwan}

\author{Jian Zhou}
\affiliation {Department of Materials Science and Engineering, Nanjing University, Nanjing 210093, China}

\author{Yugui Yao}
%\email{ygyao@bit.edu.cn}
\affiliation {School of Physics, Beijing Institute of Technology, Beijing 100081, China}

\author{Qian Niu}
\affiliation {Department of Physics, The University of Texas at Austin, Texas 78712, USA}
\affiliation {International Center for Quantum Materials and Collaborative Innovation Center
of Quantum Matter, Peking University, Beijing 100871, China}

\date{\today}

\begin{abstract}

Magneto-optical Kerr effect, normally found in magnetic materials with nonzero magnetization such as 
ferromagnets and ferrimagnets, 
%believed that originates from both the spin-orbit coupling and the ferromagnetic exchange splitting, 
has been known for more than a century. Here, using first-principles density functional theory, 
we demonstrate large magneto-optical Kerr effect in high temperature noncollinear antiferromagnets 
Mn$_{3}X$ ($X$ = Rh, Ir, or Pt), in contrast to usual wisdom. 
The calculated Kerr rotation angles are large, being %up to $\sim$0.6$^{\circ}$ and  
comparable to that of transition metal magnets such as bcc Fe.  
The large Kerr rotation angles and ellipticities are found to originate from the lifting of
the band double-degeneracy due to the absence of spatial symmetry in the Mn$_{3}X$
noncollinear antiferromagnets which together with the time-reversal symmetry would preserve the Kramers theorem. 
%be nonvanished in Mn$_{3}X$ without the ferromagnetic exchange splitting,because of the rigorous breaking 
%of the time-reversal symmetry the Kerr rotation angle and the corresponding ellipticity can definitely 
%be nonvanished in Mn$_{3}X$ without the ferromagnetic exchange splitting, in contrast to the usual recognition.  
%The Kerr rotation angle of Mn$_{3}$Pt is as large as $\sim$0.6$^{\circ}$, which is slightly larger 
%than that of ordinary transition metals.  
Our results indicate that Mn$_{3}X$ would provide a rare material platform for exploration of 
subtle magneto-optical phenomena in noncollinear magnetic materials without net magnetization.

\end{abstract}

\pacs{71.15.Rf, 75.30.-m, 75.50.Ee, 78.20.Ls}

\maketitle

\section{Introduction}

Magneto-optical coupling effects reflecting the interactions between light and magnetism are 
fundamental phenomena in solid state physics.~\cite{Antonov2004}  Originally, Faraday~\cite{Faraday1846} 
and Kerr~\cite{Kerr1877} discovered, respectively, that when a linearly polarized light beam  hits a magnetic material,
the polarization plane of the transmitted and reflected light beams rotates.
Although magneto-optical Faraday and Kerr effects have been 
known for over a century, they have become the subjects of intense investigations only in the past decades, 
mainly due to the applications of optical means in modern high-density data-storage technology.~\cite{Mansuripur1995}  
Faraday effect attracts less attention than Kerr effect because it can only occur in ultra-thin films, 
where complexities of multiple reflections and discontinuous polarizations at the interfaces with the substrate
arise.  In contrast, magneto-optical Kerr effect (MOKE) is widely used as a powerful probe 
of the electronic and magnetic properties of materials, such as the domain wall,~\cite{Jiang2013a,Jiang2013b} 
surface plasma resonance,~\cite{Kravets2005,Razdolski2013} magnetic anisotropy,~\cite{Lehnert2010,He2015} 
and topological insulator.~\cite{Tse2010,Aguilar2012}

Band exchange splitting caused by magnetization together with relativistic spin-orbit coupling (SOC) has been 
recognized as the origin of 
MOKE.~\cite{Arg55,Erskine1973a,Erskine1973b,Guo1994,Guo1995,Ravindran1999,Kim1999,Stroppa2008,Rosa2015}  
Therefore, MOKE has been explored extensively in various ferromagnetic transition metals as well as their alloys and compounds.
By ferromagnets here we meant the magnetic materials with net magnetization including ferrimagnets.
Experimentally, Erskine and Stern~\cite{Erskine1973a,Erskine1973b} first reported the MOKE spectra of bulk Fe, Co, Ni, and Gd, 
and discussed their relationships with $d$-band widths and electron spin polarizations.  After that, 
large Kerr rotation angles of $\sim$1.0$^{\circ}$ were observed in several Mn-based ferromagnetic alloys, 
such as PtMnSb,~\cite{Engen1983} MnBi,~\cite{Di1992} and MnPt$_{3}$.~\cite{Kato1995} On the theoretical side, 
first-principles density functional calculations can directly capture the MOKE spectra with an impressive accuracy 
compared to experiments.  For instance, Guo and Ebert~\cite{Guo1994,Guo1995} studied the 
the MOKE spectra in bulk Fe and Co as well as their multilayers.  Kim \textit{et al.}~\cite{Kim1999} 
focused on the surface effect and structural dependence of the MOKE spectra in Co thin films and CoPt alloys, 
and Stroppa \textit{et al.}~\cite{Stroppa2008} analyzed the electronic structure and 
magneto-optical property of uniformly Mn-doped GaAs. Very recently, Rosa \textit{et al.}~\cite{Rosa2015} 
also performed the study of the magneto-optical property of Mn-doped GaAs in a special digital 
ferromagnetic heterostructure.  Moreover, Ravindran \textit{et al.}~\cite{Ravindran1999} investigated the magnetic, 
optical, and magneto-optical properties of manganese pnictides and found a systematic increase of 
the Kerr rotation angles from MnAs, to MnSb, and to MnBi.

Although MOKE experiments have been almost always conducted on various types of ferromagnets 
in the past~\cite{Erskine1973a,Erskine1973b,Guo1994,Guo1995,Ravindran1999,Kim1999,Stroppa2008,Rosa2015,Engen1983,Di1992,Kato1995},
no explicit conclusion has been established that MOKE \textit{must be} absent when either magnetization %exchange splitting 
or SOC is not present. 
%On one hand, Stroppa \textit{et al.}~\cite{Stroppa2008} indeed mimicked a nonvanished 
%MOKE spectrum of Mn-doped GaAs using the superposition of the majority and minority spin bands (without SOC), 
%since the band splittings induced by SOC are weak compared to those by ferromagnetic exchange field.  On the other hand, 
In particular, whether MOKE can arise from a spin non-polarized system (without magnetization) 
or not is still an open question.  Remarkably, Chen \textit{et al.}~\cite{Chen2014} recently revealed that 
the anomalous Hall effect, which has a physical origin akin to that of MOKE, is large in noncollinear 
antiferromagnet Mn$_{3}$Ir with zero net magnetization.  This surprising result stems from the fact 
that in a three-sublattice kagome lattice with a noncollinear triangle antiferromagnetic configuration, 
not only the time-reversal symmetry $\mathcal{T}$ is broken but also there is no spatial symmetry
operation $\mathcal{S}$ which, in combination with $\mathcal{T}$, \textit{i.e.}, the $\mathcal{TS}$, 
is a good symmetry that preserves the Kramers theorem. In other words, band exchange splitting exists 
in this system despite of zero net magnetization. Naturally, it would
be interesting to explore possible MOKE in Mn$_{3}$Ir as well as its isostructural materials Mn$_{3}$Rh and Mn$_{3}$Pt, 
which are widely considered as the promising candidates in information-storage devices due to their 
prominent exchange-bias properties~\cite{Kohn2013} and high N\'{e}el temperatures.~\cite{Tomeno1999,Yamauchia1998,Kren1968}

In this paper, we present a comprehensive first-principles study of MOKE in spin non-polarized systems, 
focusing on antiferromagnets Mn$_{3}X$ ($X$ = Rh, Ir, or Pt) in two low-energy noncollinear spin structures.  
We show that, because of the strong SOC and the breaking of band double-degeneracy, 
the absorption rates of the left- and right-circularly polarized lights differ significantly, 
giving rise to a previously undetected MOKE in these antiferromagents without net magnetization.  The Kerr rotation angles 
of Mn$_{3}X$ increase from $X$ = Rh, to Ir, and to Pt, due to the increased SOC strength of the $X$ atom.  
The largest one of $\sim$0.6$^{\circ}$ in Mn$_{3}$Pt is comparable to %slightly larger than i
those of elemental transition metals, such as Fe and Co, reported previously~\cite{Guo1994,Guo1995}.  
Our first-principles calculations also show that the MOKE would diminish if the SOC 
is switched off, demonstrating the essiential role of the SOC for the occurrence of the MOKE in these systems. 
%Pronounced differences in the MOKE spectra between the T1 and T2 spin structures are found. 
%, which can be used to help identify the magnetic structure of Mn$_{3}$Ir.  
Our theoretical results suggest noncollinear antiferromagnets Mn$_{3}X$ to be an interesting 
material platform for further studies of novel magneto-optical phenomena and technological applications.

\section{Theory and computational details}

MOKE generally refers to the change in the polarization property of light when it interacts with magnetism, 
that is, a linearly polarized light shone on the surface of a magnetic sample will become elliptically polarized 
in the reflected beam.  The ellipticity $\epsilon_{K}$ and Kerr angle $\theta_{K}$ (rotation 
of the major axis relative to the polarization axis of the incident beam) are widely used to 
probe and characterize magnetic materials. $\epsilon_{K}$ and $\theta_{K}$ are usually combined into 
the complex Kerr angle, $\phi_{K}=\theta_{K}+i\epsilon_{K}$. 
Depending on the directions of photon propagation and magnetization vector 
with respect to the surface plane, there are three different geometries of the Kerr effect, 
namely, the polar, longitudinal, and transverse geometries.  Of these, the polar geometry usually has 
the largest complex Kerr angle and thus is the most interesting one in connection with technological 
applications.  In this paper, we consider the polar geometry as a prototype and the other two geometries 
can be obtained similarly.  

For a solid with at least threefold rotational symmetry, the elements of optical conductivity tensor 
satisfy $\sigma_{xx}=\sigma_{yy}$ and $\sigma_{xy}=-\sigma_{yx}$.  In such a case, the absorptive parts 
of optical conductivity tensor (real diagonal and imaginary off-diagonal elements) due to interband
transitions, can be obtained using 
the Kubo's formula within the linear response theory,~\cite{Kubo1957,Wang1974,Callaway1991}
\begin{eqnarray}
\sigma_{xx}^{1}\left(\omega\right) & = & \frac{\lambda}{\omega}\sum_{\bm{k},jj^{\prime}}\left[\left|\Pi_{jj^{\prime}}^{+}\right|^{2}+\left|\Pi_{jj^{\prime}}^{-}\right|^{2}\right]\delta\left(\omega-\omega_{jj^{\prime}}\right),\label{eq:1}
\\
\sigma_{xy}^{2}\left(\omega\right) & = & \frac{\lambda}{\omega}\sum_{\bm{k},jj^{\prime}}\left[\left|\Pi_{jj^{\prime}}^{+}\right|^{2}-\left|\Pi_{jj^{\prime}}^{-}\right|^{2}\right]\delta\left(\omega-\omega_{jj^{\prime}}\right),\label{eq:2}
\end{eqnarray}
where $\lambda=\frac{\pi e^{2}}{2\hbar m^{2}V}$ is a material specific constant, $\hbar\omega$ the photon energy, 
$\hbar\omega_{jj^{\prime}}$ the energy difference between the occupied and unoccupied bands at the same $\bm{k}$-point, 
and $\Pi_{jj^{\prime}}^{\pm}=\langle {\bf k}j|\frac{1}{\surd{2}}\left(\hat{p_x}\pm i\hat{p_y}\right)|{\bf k}j'\rangle$ 
the dipole matrix elements for circularly polarized light with $+$ and $-$ helicity, respectively. 
%interband transition operator defined by the momentum operators.  
The corresponding dispersive parts, namely, the imaginary diagonal element $\sigma^{2}_{xx}\left(\omega\right)$ 
and real off-diagonal element $\sigma^{1}_{xy}\left(\omega\right)$, can be obtained by use 
of the Kramer-Kronig transformation.~\cite{Bennett1965}  
%Both the absorptive and dispersive parts 
%are broadened with a Lorentzian of 0.1 eV to simulate the finite lifetime effect of electron.

In the polar geometry, the complex Kerr angle of a sample with higher than threefold rotational symmetry, 
is simply given as,~\cite{Kahn1969}
\begin{equation}\label{eq:3}
\theta_{K}+i\epsilon_{K}=\frac{-\sigma_{xy}}{\sigma_{xx}\sqrt{1+i\left(4\pi/\omega\right)\sigma_{xx}}},
\end{equation}
which can be explicitly evaluated from the optical conductivity tensor calculated from the electronic structure 
of the solid concerned. Since the intraband 
transitions contribute little to the off-diagonal elements of optical 
conductivity tensor in the magnetically ordered materials~\cite{Antonov2004} and affect mainly 
the MOKE spectra below 1$\sim$2 eV,~\cite{Guo1994,Guo1995,Stroppa2008} here we take into account only 
the interband transition contribution as expressed in Eqs. (\ref{eq:1}--\ref{eq:3}).

\begin{figure}
\includegraphics[width=\columnwidth]{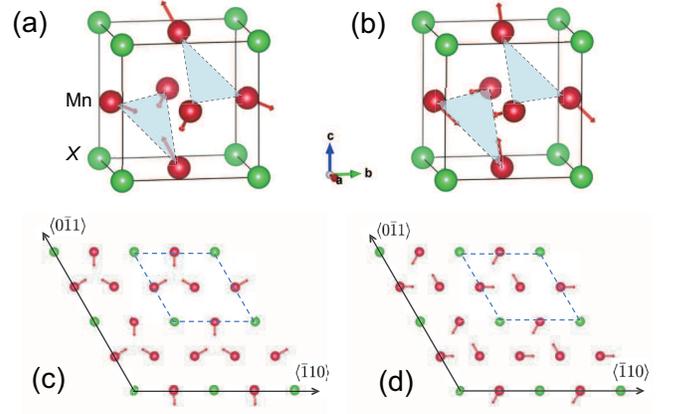}
\caption{(Color online) Cubic $L1_2$ crystal structure of Mn$_{3}X$ ($X$ = Rh, Ir, or Pt) 
with T1 (a) and T2 (b) spin configurations, as well as the corresponding (111) planes in (c) and (d), 
respectively.  The red and green balls represent Mn and $X$ atoms, respectively.  
The dashed lines in (c) and (d) stand for the primitive cell of the kagome lattice.}
\label{fig:crystal}
\end{figure}

In this paper, we consider ordered cubic $L1_2$ Mn$_{3}$Rh, Mn$_{3}$Ir, and Mn$_{3}$Pt alloys and
adopt the experimental lattice constants of 3.813 \AA,~\cite{Kren1968} 3.785 \AA,~\cite{Yamaoka1974} 
and 3.833,\AA~\cite{Kren1968}, respectively. 
The total energy and electronic structure are calculated based on first-principles density functional theory 
with the generalized-gradient approximation in the form of Perdew-Berke-Ernzerhof~\cite{Perdew1996}. 
%to the exchange-correlation functional in the Perdew-Berke-Ernzerhof's form~\cite{Perdew1996}. 
The accurate frozen-core full-potential projector-augmented wave method~\cite{Blochl1994}, 
as implemented in the Vienna \textit{ab initio} simulation package (\textsc{vasp})~\cite{Kresse1993,Kresse1996}, is used. 
% with projector-augmented wave method~\cite{Blochl1994} and exchange-correlation functional in the 
%Perdew-Berke- Ernzerhof's form within the generalized-gradient approximation.~\cite{Perdew1996}  
The fully relativistic projector-augmented potentials are adopted in order to include the SOC.
The valence configurations of Mn, Rh, Ir, and Pt atoms taken into account in the calculations 
are $3d^{6}4s^{1}$, $4d^{8}5s^{1}$, $5d^{8}6s^{1}$, and $5d^{9}6s^{1}$, respectively.  
A large plane-wave energy cutoff of 350 eV and a fine Monkhorst-Pack $\bm{k}$-point mesh 
of 16$\times$16$\times$16 are used for the self-consistent electronic structure calculations.  
%To obtain the coplanar triangular spin texture, the directions of spin magnetic moment 
%are fixed forcibly in the (111) plane by adding a penalty functional to total effective potential.  
For the calculation of optical conductivity tensors, a denser $\bm{k}$-point mesh of 20$\times$20$\times$26 
(8833 $k$-points in the irreducible Brillouin zone) is used in the tetrahedron integration.

\begin{table}[b]
\caption{The calculated total energies and spin magnetic moments for the T1, T2 and T3 magnetic structures 
of the Mn$_{3}X$ alloys. Small total spin magnetic moments, being parallel to the $\left\langle111\right\rangle$ direction 
and due to the spin-canting caused by SOC, exist. The $X$ atoms have a zero magnetic moment, 
dictated by the site-symmetry of their positions in the crystal structure. 
The available previously reported Mn spin moments are also listed
for comparison. Note that the T2 structure reported in Ref. \onlinecite{Kubler1988} is named the T3 structure here 
because it has a higher energy than the T2 structure here. The listed total energies are relative to that of the T1 state.}
\label{tab:spin}
\begin{ruledtabular}
\begin{tabular}{ccccc}
\multicolumn{1}{c}{} & 
\multicolumn{1}{c}{} & 
\multicolumn{1}{c}{Energy} &
\multicolumn{1}{c}{$m_{\textrm{Mn}}$} &
%\multicolumn{1}{c}{$m^{s}_{\textrm{Mn}}$ (others) } &
\multicolumn{1}{c}{$m_{\textrm{tot}}$ ($\left\langle 111\right\rangle _{\parallel}$)} \\
			
\multicolumn{1}{c}{} & 
\multicolumn{1}{c}{} & 
\multicolumn{1}{c}{(meV)} &
\multicolumn{1}{c}{($\mu_{B}/$atom)} &
%\multicolumn{1}{c}{($\mu_{B}/$atom)} &
\multicolumn{1}{c}{($\mu_{B}/$cell)} \\
			
			\hline
Mn$_{3}$Rh  & T1  &  0.0  &  3.17, 3.6\footnotemark[1], 3.10\footnotemark[2], 2.78\footnotemark[3]   & -0.001  \\
            & T2  &  0.35 &  3.18                         &  0.002  \\
            & T3  &  1.33 &  3.19, 3.10\footnotemark[2]   &  0.000  \\
Mn$_{3}$Ir  & T1  &  0.0  &  2.96,  2.91\footnotemark[4], 2.66\footnotemark[5], 2.62\footnotemark[6] & -0.029  \\
            & T2  &  2.06 &  2.96  &  0.027  \\
            & T3  &  8.47 &  2.97  &  0.000  \\
Mn$_{3}$Pt  & T1  &  0.0  &  3.12,  3.0\footnotemark[1], 2.93\footnotemark[2]  & -0.013  \\
	    & T2  &  0.76 &  3.11                         &  0.012  \\
            & T3  &  2.92 &  3.15  2.93\footnotemark[2]   &  0.000  \\
\end{tabular}
\end{ruledtabular}
\footnotemark[1]{Ref.~\onlinecite{Kren1968} (experiment),}
\footnotemark[2]{Ref.~\onlinecite{Kubler1988} (theory),}
\footnotemark[3]{Ref.~\onlinecite{Sakuma2002} (theory),}
\footnotemark[4]{Ref.~\onlinecite{Chen2014} (theory),}
\footnotemark[5]{Ref.~\onlinecite{Szunyogh2009} (theory),}
\footnotemark[6]{Ref.~\onlinecite{Sakuma2003} (theory).}
\end{table}

%\begin{figure}
%\includegraphics[width=\columnwidth]{Mn3XFig2.eps}
%\caption{(Color online) The real diagonal (a)(b) and imaginary off-diagonal (c)(d) components of optical conductivity tensor for T1 and T2 spin configurations of the Mn$_{3}X$ alloys.}
%\label{fig:adsorption}
%\end{figure}

\section{Results and Discussion}

In this section, we first present the calculated total energy and magnetic properties of the low energy noncollinear 
magnetic structures (T1, T2 and T3) of Mn$_{3}X$ ($X$ = Rh, Ir, or Pt) and also compared our results with 
available previous reports in Sec. III A. Then, the calculated optical conductivity, the key ingredient 
for evaluating the MOKE, for the two low energy spin structures (T1 and T2) is reported in Sec. III B.  
Finally, we present the large magneto-optical Ker effect in the ordered Mn$_{3}X$ alloys in Sec. III C.

\subsection{Magnetic structure}

There are two kinds of crystal structures for Mn$_{3}X$ alloys, namely, ordered $L1_{2}$-type and disordered $\gamma$-phase.  
The ordered alloys were found to be noncollinear antiferromagnetic with one of the two nearly degenerate spin configurations,
namely, the T1 and T2 triangle structures, as shown in Figs.~\ref{fig:crystal}(a) and ~\ref{fig:crystal}(b), respectively.  
The Mn spin magnetic moments basically lie in the (111) plane and point to the center (along the edge) of the triangle, 
forming three nearest-neighboring Mn sublattices, for the T1 (T2) configuration, which can be viewed as two-dimensional 
kagome lattices shown in Figs.~\ref{fig:crystal}(c) and ~\ref{fig:crystal}(d).  Due to strong exchange-interactions 
acting on the Mn moments, the N\'{e}el temperatures in these Mn-based alloys is as high 
as 475 K in Mn$_{3}$Pt,~\cite{Kren1968} 855 K in Mn$_{3}$Rh,~\cite{Yamauchia1998} and 960 K in Mn$_{3}$Ir.~\cite{Tomeno1999}

The calculated total energy and spin magnetic moments together with previously reported experimental and theoretical results 
are listed in Table~\ref{tab:spin}. Clearly, T1 is the magnetic ground state of the Mn$_{3}X$ alloys, being in good agreement
with previous experimental~\cite{Kren1968} and theoretical~\cite{Kubler1988} works. Nevertheless, the energy difference 
between the T1 and T2 spin configurations is small (being in the order of $\sim$1 meV), $i.e.$, the T1 and T2 are
nearly degenerate. Indeed, in the absence of the SOC, all the T1, T2 and T3 spin structures have the same total energy because they
are equivalent. Note that the T2 configuration here is not the same as the T2 magnetic structure reported in 
Ref. ~\onlinecite{Kubler1988}. Our total energy calculations show that for Mn$_{3}$Ir, the total energy of the T2 structure in 
Ref. ~\onlinecite{Kubler1988} is $\sim$6.4 meV higher than the T2 configuration here, and thus should be refered to as
the T3 spin configuration. 

The calculated spin magnetic moment of the $X$ atom is always zero, due to its special site-symmetry, 
while those of the Mn atoms have nearly identical values of $\sim3\mu_{B}$ for all the three states.  
The calculated spin magnetic moments agree fairly well with previous 
reports.~\cite{Chen2014,Kren1968,Kubler1988,Sakuma2002,Szunyogh2009,Sakuma2003}   
Further inspecting the total magnetization, we find a nonvanishing component 
along the $\left\langle\bar{1}\bar{1}\bar{1}\right\rangle$ ($\left\langle111\right\rangle$) 
direction for T1 (T2) states, because the Mn moments rotate slightly away from the (111) plane within  
a very small angle of $\sim0.1^{\circ}$.  
%The nonzero net magnetization in Mn$_{3}X$ alloys were ignored in all previous works, except for Mn$_{3}$Ir in Ref.~\onlinecite{Chen2014}.  
Nonetheless, the Mn$_{3}X$ alloys could still be safely considered as spin non-polarized systems in the sense that the net total 
magnetization is very small and hardly affacts the physical quantities of interest here, such as optical conductivities and MOKE spectra, 
as will be discussed in the next subsection. Note that this small spin-canting is caused by the presence of the SOC.
Interestingly, such a small spin-canting does not occur in the T3 structure (Table I).
%are almost identical to the absolutely non-polarized case obtained by forcibly fixing the Mn moments lying in the (111) plane.

\begin{figure}
\includegraphics[width=\columnwidth]{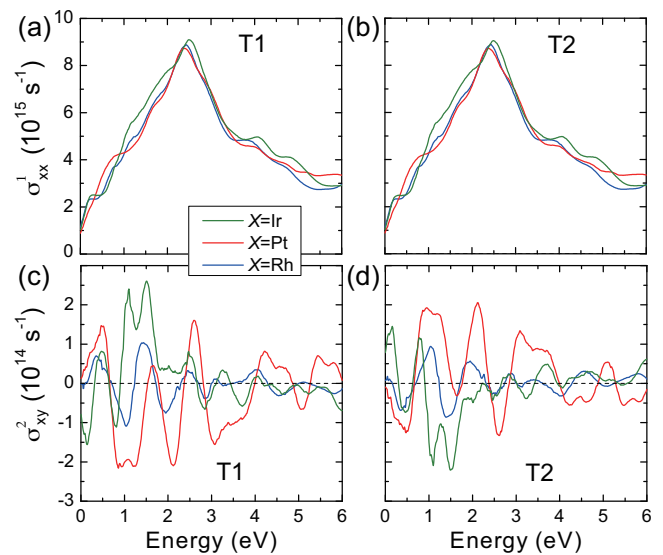}
\caption{(Color online) (a) and (b) The calculated real diagonal ($\sigma^{1}_{xx}$) as well as (c) and (d) 
imaginary off-diagonal ($\sigma^{2}_{xy}$) components 
of the optical conductivity tensor for the T1 and T2 spin configurations of the Mn$_{3}X$ alloys.
Both $\sigma^{1}_{xx}$ and $\sigma^{2}_{xy}$ are broadened with a Lorentzian of 0.1 eV 
to simulate the finite lifetime effect of electron.}
\label{fig:adsorption}
\end{figure}

\subsection{Optical conductivity}

The adsorptive parts of optical conductivity, \textit{i.e.}, $\sigma^{1}_{xx}$ and $\sigma^{2}_{xy}$, have direct physical interpretations.  
From Eqs. (\ref{eq:1}) and (\ref{eq:2}), it is clear that $\sigma^{1}_{xx}$ measures the average in the absorption 
of left- and right-circularly polarized light while $\sigma^{2}_{xy}$ measures the corresponding difference.  
In Figs.~\ref{fig:adsorption}(a) and~\ref{fig:adsorption}(b), we show the $\sigma^{1}_{xx}$ for the T1 and T2 spin configurations 
in the energy range of 0$\sim$6 eV, respectively. %Both the absorptive and dispersive parts
%are broadened with a Lorentzian of 0.1 eV to simulate the finite lifetime effect of electron. 
Since the $\sigma^{1}_{xx}$ is directly related to the joint density of states 
and interband transition probability but does not depend strongly on the details of the spin structure,~\cite{Stroppa2008} 
it is not surprising that the calculated $\sigma^{1}_{xx}$ are nearly the same for the two different spin configurations.  
Moreover, the $\sigma^{1}_{xx}$ for all 
the three Mn$_{3}X$ alloys have similar behaviors, mainly due to their isostructural nature. In particular, the $\sigma^{1}_{xx}$ 
for all the three alloys has a prominent broad peak centered at 2.5 eV.  The $\sigma^{2}_{xy}$ for the T1 and T2 configurations are displayed 
in Figs.~\ref{fig:adsorption}(c) and~\ref{fig:adsorption}(d), respectively. For both configurations, the $\sigma^{2}_{xy}$ of all 
the three Mn$_{3}X$ alloys have pronounced oscillatory peaks in the low energy region and its magnitude reduces gradually to 
a small value above 6 eV (not shown). For each individual Mn$_{3}X$ alloy, the $\sigma^{2}_{xy}$ for the T1 and T2 states differ in sign, 
due to the opposite total spin magnetic moments (see Table.~\ref{tab:spin}), although they are similar in line shape and magnitude.  
Positive (negative) $\sigma^{2}_{xy}$ suggests that the interband transitions are dominated by the excitations due to
the left (right) circularly polarized light. Interestingly, the sign of the $\sigma^{2}_{xy}$ for the T1 structure
can be reversed by reversing the Mn spin moments while that for the T2 structure remains unchanged when the chirality
of the spin structure is reversed.

\begin{figure}
\includegraphics[width=\columnwidth]{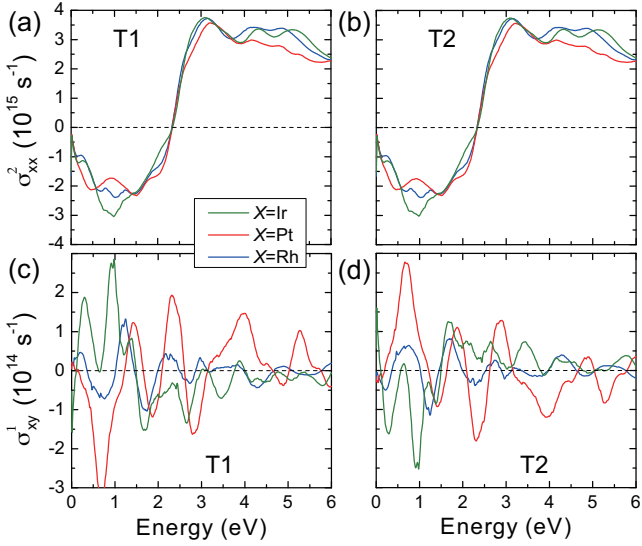}
\caption{(Color online) (a) and (b) The calculated imaginary diagonal ($\sigma^{2}_{xx}$) as well as (c) and (d) 
real off-diagonal ($\sigma^{1}_{xy}$) components of optical conductivity tensor for the T1 and T2 spin configurations 
of the Mn$_{3}X$ alloys.  Both $\sigma^{2}_{xx}$ and $\sigma^{1}_{xy}$ are broadened with a Lorentzian of 0.1 eV 
to simulate the finite lifetime effect of electron.}
\label{fig:dispersion}
\end{figure}

Physically speaking, the \textit{dc} limit of the imaginary off-diagonal element of optical 
conductivity, $\sigma^{2}_{xy}\left(\omega=0\right)$, is nothing but the anomalous Hall conductivity,~\cite{Guo2015,Xiao2010} 
which can also be precisely evaluated by the integration of the Berry curvature over the Brillouin zone.~\cite{Yao2004,Guo2014}  
Chen \textit{et al.}~\cite{Chen2014} recently pointed out that the anomalous Hall effect can arise from 
noncollinear antiferromagnet Mn$_{3}$Ir in the T1 spin structure without net magnetization due to 
the absence of certain spatial symmetries. This could be understood in terms of the fact 
that in the kagome lattice [the (111) plane of the Mn$_{3}X$ alloys, as shown in Figs.~\ref{fig:crystal} 
(c) and (d)] there is no spatial symmetry $\mathcal{S}$ such as mirror and rotation that in combination 
with the time-reversal symmetry $\mathcal{T}$ (i.e., the $\mathcal{ST}$) can be a good symmetry such that 
the band Kramers degeneracy will be kept in the system with broken time-reversal symmetry (TRS). 
This is certainly in contrast to the case of, e.g., a collinear bipartite antiferromagnet, where
the combination of the translational operation of half of a lattice vector with the time-reversal operation 
is a good symmetry that will preserve the band Kramers degeneracy despite of the broken TRS due to 
the antiferromagnetism. 
%Jby combining the translational operation (half of the lattice vector) and the time-reversal operation.  
%From this symmetry point of view, Mn$_{3}$Ir and also Mn$_{3}$Rh and Mn$_{3}$Pt, would \textit{rigorously} break the time-reversal symmetry.  
This lifting of the band Kramers degeneracy together with the strong SOC in the Mn$_{3}X$ alloys gives rise to the nonzero anomalous Hall conductivity.  
Similarly, one can expect that the $\sigma_{xy}$ at optical frequencies would be nonzero as well and from Eq. (\ref{eq:3}), 
result in nontrivial magneto-optical Kerr effect in the Mn$_{3}X$ alloys, which will be discussed in next subsection.
Of course, one may argue that the nonzero $\sigma^{2}_{xy}$ could be due to the nonzero total spin magnetic moment
in the T1 and T2 spin structures (Table I). To clarify this, we also calculate the $\sigma^{2}_{xy}$ spectrum
from the electronic structure with a zero spin magnetic moment obtained by forcing all the Mn moments lying
in the (111) plane, and the calculated $\sigma^{2}_{xy}$ spectrum (not shown here) is nearly identical
to that obtained without fixing the Mn moments to lie in the (111) plane.

The dispersive parts of optical conductivity, \textit{i.e.}, $\sigma^{2}_{xx}$ and $\sigma^{1}_{xy}$, can be obtained 
from the corresponding absorptive parts by use of the Kramers-Kronig transformation.  In Fig.~\ref{fig:dispersion}, 
we plot the $\sigma^{2}_{xx}$ and $\sigma^{1}_{xy}$ for the T1 and T2 spin configurations, respectively.  
Figs.~\ref{fig:dispersion}(a) and~\ref{fig:dispersion}(b) show that, similar to the $\sigma^{1}_{xx}$, 
the $\sigma^{2}_{xx}$ are almost the same for the T1 and T2 configurations. The $\sigma^{2}_{xx}$ for 
all the Mn$_{3}X$ alloys have common characteristics such as a broad valley around 1.0$\sim$1.5 eV, 
a negative to positive crossing point at 2.5 eV, and a broad plateau above 3.0 eV.  
Figures~\ref{fig:dispersion}(c) and~\ref{fig:dispersion}(d) show that the $\sigma^{1}_{xy}$ for the T1 and T2 
configurations have similar profiles but opposite in sign, and gradually decay to small values in the high 
energy region, which are similar to the behavior of $\sigma^{2}_{xy}$.

\begin{figure}
\includegraphics[width=\columnwidth]{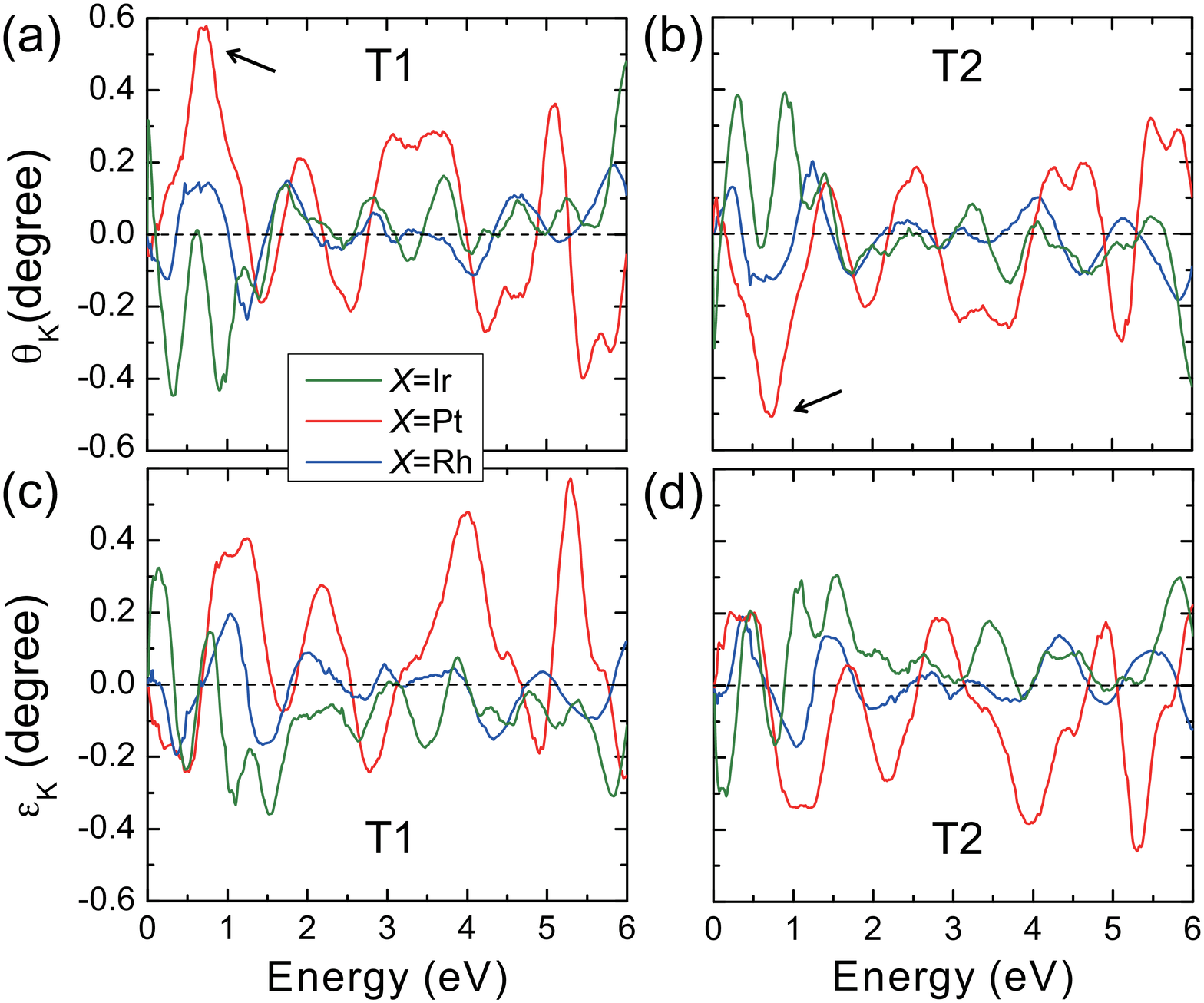}
\caption{(Color online)  The calculated complex Kerr angles for the T1 (left panels) and T2 (right panels)
spin configurations of the Mn$_{3}X$ alloys: Kerr rotations (upper panels) and Kerr ellipticities
(lower panels). The arrows indicate the largest Kerr rotation angles.}
\label{fig:MOKE}
\end{figure}

\subsection{Magneto-optical Kerr effect}

After discussing the magnetic and optical properties of the Mn$_{3}X$ alloys, we now turn our attention 
to their magneto-optical property.  From the complex Kerr angle spectra presented in Fig.~\ref{fig:MOKE}, 
one can find their key features as follows:  (1) The calculated Kerr rotation angles ($\theta_{K}$) 
and ellipticities ($\epsilon_{K}$) for the T1 and T2 states have opposite signs, inheriting from 
the behaviors of the off-diagonal elements of optical conductivity, $\sigma^{1}_{xy}$ [see  Figs.~\ref{fig:dispersion}(c) 
and~\ref{fig:dispersion}(d)] and $\sigma^{2}_{xy}$ [see Figs.~\ref{fig:adsorption}(c) and~\ref{fig:adsorption}(d)]; 
(2) The sign reversals of $\theta_{K}$ and $\epsilon_{K}$ are frequent in the given energy range.  
When $\epsilon_{K}$ crosses the zero line, a peak turns up in the corresponding $\theta_{K}$ spectrum, and vice versa, 
which may be ascribed to the Kramers-Kronig relation; (3) The overall features and maximum values of $\theta_{K}$ 
and $\epsilon_{K}$ have a size sequence of Mn$_{3}$Pt $>$ Mn$_{3}$Ir $>$ Mn$_{3}$Rh, which suggests that 
the SOC strength of the $X$ atom contributes significantly to the enhancement of the MOKE spectra; 
(4) The largest Kerr rotation angle that appears in Mn$_{3}$Pt is $\sim$0.6$^{\circ}$ at incident 
photon energy of 0.7 eV. This angle arising from a noncollinear antiferromagnet is remarkably large and 
comparable to those of transition metals, such as the bulk and multilayers of Fe and Co studied 
earlier.~\cite{Guo1994,Guo1995}

Finally, we attempt to analyze the origin of the large MOKE in the Mn$_{3}X$ alloys. Equation (\ref{eq:3})
indicates that a peak in a Kerr spectrum could stem from either a small $\sigma_{xx}$ in the denominator 
or a large $\sigma_{xy}$ in the numerator, which are called the ``optical" and ``magneto-optical" origins, 
respectively. From Figs.~\ref{fig:dispersion}(c) and~\ref{fig:MOKE}(a), one can observe that the positions 
of the peaks of $\sigma^{1}_{xy}$ and $\theta_{K}$ for the T1 state overlap with each other, 
such as the first peak at 0.7 eV in Mn$_{3}$Pt and the twin peaks in the low energy range of 0$\sim$1.0 eV in Mn$_{3}$Ir, 
and the same can be seen for the T2 state by comparing Fig.~\ref{fig:dispersion}(d) with Fig.~\ref{fig:MOKE}(b).  
Furthermore, the positions of the peaks of $\epsilon_{K}$ and $\sigma^{2}_{xy}$ are close,  
as shown in Figs.~\ref{fig:adsorption}(c) and~\ref{fig:adsorption}(d) as well as Figs.~\ref{fig:MOKE}(c) 
and~\ref{fig:MOKE}(d), respectively. Thus, the magnitudes of the peaks of $\theta_{K}$ and $\epsilon_{K}$ 
are modulated by the $\sigma_{xx}$, as shown in Figs.~\ref{fig:adsorption}(a) and~\ref{fig:adsorption}(b) 
as well as Figs.~\ref{fig:dispersion}(a) and~\ref{fig:dispersion}(b), respectively.  Since the Kerr rotation 
angle and also the ellipticity entangle in a complicated way with both the real and imaginary components of 
the optical conductivity tensor, there are no analytic forms for strictly separating the ``optical" 
and ``magneto-optical" origins for them. On the other hand, the nonzero $\sigma_{xy}$ is clearly the root cause 
for the emergence of the Kerr effect in this kind of noncollinear antiferromagnets, 
as already discussed in the preceding subsection. This is corroborated by our test calculations
which show that both the $\sigma_{xy}$ and MOKE in these antiferromagnets would become zero 
without the SOC included. Therefore, it can be concluded that the large MOKE 
in the Mn$_{3}X$ alloys has a ``magneto-optical" origin rather than the ``optical" origin.

\section{Summary}

In conclusion, using first-principles density functional calculations, we have investigated the possible 
magneto-optical Kerr effect in noncollinear antiferromagnets Mn$_{3}$Rh, Mn$_{3}$Ir and Mn$_{3}$Pt.  
We found that the Kerr rotation angle can be as large as $\sim$0.6$^{\circ}$ in Mn$_{3}$Pt, which 
is comparable to that in elemental transition metal ferromagnets such as bcc Fe.  We also discussed 
the differences in the magneto-optical responses for the T1 and T2 spin configurations.  The surprisingly 
large magneto-optical Kerr effect in noncollinear antiferromagnets with nearly zero magnetization 
is attributed to the nontrivial off-diagonal components of optical conductivity, \textit{i.e.,} 
having the so-called ``magneto-optical" origin.  Our results demonstrate that one cannot
assume {\it a priori} vanishing magneto-optical Kerr effect in antiferromagnets with zero net magnetization.
The large Kerr rotation angle, plus other interesting physical properties 
of the Mn$_{3}X$ alloys, such as prominent exchange-bias properties~\cite{Kohn2013} and 
high N\'{e}el temperatures~\cite{Tomeno1999,Yamauchia1998,Kren1968}, would make 
these materials an exciting platform for exploring novel information-storage devices.

\begin{acknowledgments}
Q.N. and G.Y.G. thank Hua Chen and Alan MacDonald for stimulating discussions.
W.F. and G.Y.G acknowledge support from the Ministry of Science and Technology, the Academia Sinica and NCTS of Taiwan.
W.F. and Y.Y. were supported in part by the MOST Project of China (Grant Nos. 2014CB920903 and 2011CBA00108), 
the NSF of China (Grant Nos. 11374033, 11225418, and 11174337), the Specialized Research Fund 
for the Doctoral Program of Higher Education of China (Grant Nos. 20121101110046 and 20131101120052), 
and the Basic Research Fund of Beijing Institute of Technology (Grant No. 20141842004). 
Q.N. was supported in part by DOE-DMSE (Grant No. DE-FG03-02ER45958) and the Welch Foundation (Grant No. F-1255).
%W.F. thanks the National Taiwan University and Academia Sinica for the hospitality during his stay in Taiwan.
W.F. also acknowledges the use of the computational resources provided by the National Supercomputer 
Center in Tianjin (NSCC--TJ).

\end{acknowledgments}

\end{document}